\documentclass{article}
 
\usepackage{graphicx}
\usepackage{amsthm,amsmath}
\usepackage[utf8]{inputenc} 
\usepackage[hidelinks]{hyperref}
\usepackage{mathtools}
\usepackage{amssymb}
\usepackage{bm}
\usepackage{fontawesome}
\usepackage{color}
\usepackage[margin=1.0in]{geometry}
\usepackage{authblk}

\newcommand{\x}{\textbf{x}}
\newcommand{\X}{\textbf{X}}
\newcommand{\uu}{\textbf{u}}
\newcommand{\UU}{\textbf{U}}

\newcommand{\f}{\textbf{f}}
\newcommand{\btheta}{\bm{\theta}}

\def\secref#1{Section \ref{#1}}

\newtheorem{thm}{Theorem}[section]

\newtheorem{remark}[thm]{Remark}

\DeclareMathOperator*{\argmin}{arg\,min}

\title{Efficient Long-Term Structural Reliability Estimation with Non-Gaussian Stochastic Models: A Design of Experiments Approach}

\date{\today}
\author[1]{Sebastian Winter}
\author[1]{Christian Agrell}
\author[1]{Juan Camilo Guevara Gómez}
\author[1]{Erik Vanem}

\affil[1]{DNV Group Research and Development}

\begin{document}

\maketitle

\begin{abstract}
Extreme response assessment is important in the design and operation of engineering structures, and is a crucial part of structural risk and reliability analyses. Structures should be designed in a way that enables them to withstand the environmental loads they are expected to experience over their lifetime, without designs being unnecessarily conservative and costly. An accurate risk estimate is essential but difficult to obtain because the long-term behaviour of a structure is typically too complex to calculate analytically or with brute force Monte Carlo simulation. Therefore, approximation methods are required to estimate the extreme response using only a limited number of short-term conditional response calculations. Combining surrogate models with Design of Experiments is an approximation approach that has gained popularity due to its ability to account for both long-term environment variability and short-term response variability. In this paper, we propose a method for estimating the extreme response of black-box, stochastic models with heteroscedastic non-Gaussian noise. We present a mathematically founded extreme response estimation process that enables Design of Experiment approaches that are prohibitively expensive with surrogate Monte Carlo. The theory leads us to speculate this method can robustly produce more confident extreme response estimates, and is suitable for a variety of domains. While this needs to be further validated empirically, the method offers a promising tool for reducing the uncertainty decision-makers face, allowing them to make better informed choices and create more optimal structures.
\end{abstract}

\section{Introduction and background}

Structural reliability assessment is an integral part of structural design and is carried out to ensure that engineering structures can withstand the environmental loads they can be expected to experience within their lifetimes. A wide range of structural responses needs to be considered, and each response is associated with a limit state that describes when the structure is expected to fail. An ultimate limit state (ULS) describes failure due to extreme loads. In probabilistic design, one will typically design against a target reliability level, corresponding to a maximum probability of failure. These failure probabilities are typically very small and associated with rare events. Hence, the uncertainties are large, and it is difficult to accurately estimate the long-term extreme responses associated with long return periods.

Structural reliability analysis is carried out to determine the reliability ($R$) or failure probability ($P_f$) of a given structural design. The limit state function (also called the performance function) $g(\boldsymbol{x})$ defines when a structure will fail. Consider a limit state of the form $g(\boldsymbol{x}) = y_{\text{capacity}} - y(\boldsymbol{x})$, where $y_{\text{capacity}}$ is the structural strength or capacity, $y(\boldsymbol{x})$ is the response (actual load or structural behaviour) on the structure and $\boldsymbol{x}$ is a vector of relevant input variables. $y(\boldsymbol{x})$ is often calculated using complex computational models, which we refer to as simulators. If the structural capacity is greater than the loads, i.e. when $g(\boldsymbol{x}) > 0$, the structure will survive. Then the reliability can in principle be calculated by integrating the probability density function of the random input variables, $\boldsymbol{X}$, over the safe region:
\begin{align}
 R = 1 - P_f = P\left[ g(\boldsymbol{X}) > 0 \right] = \int\limits_{g(\boldsymbol{X}) > 0} f_{\boldsymbol{X}}(\boldsymbol{x}) d\boldsymbol{x},
\end{align}
where $f_{\boldsymbol{X}}(\boldsymbol{x})$ denotes the joint density function of the input variables. Note, the above formulation is for deterministic performance functions.

However, in practice these integrals will often be difficult to solve analytically because both the joint environmental density function and the performance function can be complicated \cite{WGSMVanem:compareULS24}. First-Order Reliability Methods (FORM) and the Second-Order Reliability Methods (SORM) are two approximate approaches that solve the integral by approximating the failure boundary using a first- or second-order Taylor expansion, respectively, at the design point \cite{WGSMVanem:compareULS24}; see \cite{4308ecd9acab433ea0ebfe09f8dfd277} for details.

Environmental conditions are often modeled as a piece-wise stationary process, where each piece (refereed to as the short-term environment) is parametrised by a set of long-term environmental conditions. To describe the structural response, the long-term environmental conditions are used to generate short-term environment behaviour, which is used in the conditional short-term response model \cite{WGSMVanem:compareULS24}. Understanding the long-term extreme response involves integrating the conditional short-term structural response over all long-term environmental conditions, see e.g. \cite{SND:LongTermResp11, FIG17_LongTermIFORM}. While this could theoretically be achieved through brute-force Monte Carlo simulations, the computational cost of evaluating short-term responses for each environmental condition is typically prohibitive in engineering applications \cite{WGSMVanem:compareULS24}. In order to address this fundamental challenge, approximate methods are needed. Essentially two different approaches can be taken. The first is to enable more efficient short-term response analysis, for example by establishing inexpensive surrogate models. Secondly, one may establish methods that can obtain long-term extreme response estimates from a limited set of short-term response calculations. This paper presents a method that exploits both these principles by utilizing a surrogate model to emulate computationally expensive response simulations and Design of Experiments (DOE) to carefully select a limited number of environmental conditions for which to perform short-term response calculations.

The environmental contour approach is an approximate method for extreme structural response estimation \cite{WitUCBH93, HW:ENvContLin09}. It essentially solves the inverse reliability problem and is based on the assumption that the largest response occurs in the most extreme environmental conditions. Contours are constructed in the environmental variables space to correspond to extreme conditions with a specified exceedance probability, and it is assumed that the failure probability is lower than this exceedance probability if a design can withstand all conditions along the corresponding contours. Conversely, under the assumption that the largest response occurs in the most severe environmental conditions, the extreme long-term response may be estimated from a limited set of short-term response analyses selected from environmental conditions along the contours. This approach ignores the impact of short-term variability in the response. If short-term variability is expected to contribute to the largest response, the contours can be inflated or a higher quantile of the conditional short-term response can be used to provide a margin for error \cite{DNVRP-C205_21}.

Environmental contours are well established in the industry, especially for ocean engineering structures \cite{DNVRP-C205_21}. It remains an active area of research and several methods to construct such contours exist. Initial methods were based on the inverse first order reliability method (IFORM) \cite{WitUCBH93, HW:ENvContLin09}, and an alternative approach based on direct sampling in physical parameter space was proposed in \cite{Vanem:EnvCont12, Vanem:EnvCont14, HusebyConvexEnvCont21}. The direct sampling contours can also be constructed based on Voronoi cells \cite{HAVanem:ContVoronoi22}, and it can be extended to higher dimensions \cite{Vanem3Dcontour17}. Other contour methods have been proposed in the literature, see e.g. \cite{M-IH-Z:EnvContCop15, DH:NewEnvCont:OMAE2019-95993, HOWT:HighDensityContour2017, CL:ISORMContours18, ECSADESjointPaper19}, and several studies have compared various contour methods \cite{Leira:StocProcCont08, VanemB-G:JOMAE15, VanemContourCompStruc17, MNCCE-GM:AltApproachEnvCont18, ECSADESShipCase19, HetalBenchmarkREsults21, ECSADESjointPaper19}. Several studies have also discussed the uncertainties associated with environmental contours for extreme response assessment, see e.g. \cite{S-GV-HSC:UncertEnvCont15, VanemGB-G:ContourUncert18, RTGS:UncertAssHydrodynrespoWT20, KHB-P:ExtRespMonoOWTCont24}, and some recent applications of environmental contours for ultimate limit state design are reported in \cite{FIG18_LongTErmPontoon, CFO:EnvContWindBridge22, WM:LoadAnalULSFloatColumn24}.   

The environmental contour method does not implicitly account for the short-term variability of the conditional response in an environmental condition, and this leads to uncertainties and biases in extreme response estimates. Although there are ways to account for this, a DOE approach was proposed in \cite{Gramstad:2020:Seq}, see also \cite{MS:SeqSampExEvent18, Agrell:2020:DOE_SRA}, that does not rely on the assumption that the long-term extreme response is dominated by the long-term variability of environmental conditions. This provides a more robust approach that implicitly accounts for both short-term and long-term variability and that addresses some of the shortcomings with the environmental contour method. A recent case study comparing environmental contour methods with the DOE approach demonstrated that the computational efficiency is comparable, but that the DOE approach might be preferable in cases where short-term variability is important \cite{WGSMVanem:compareULS24}. 

Surrogate modelling with DOE has become an increasingly popular tool for structural reliability analysis, with a large body of work being inspired by the seminal paper \cite{EGL:AK-MCS11}. A review of the many subsequent developments is presented in \cite{TNO:AdaptMetaModRelRev21}, and an overarching framework for categorising methods is presented in \cite{MMS:ActLearnStrucRel22}. While much work has been done in the field, the aforementioned papers (and the works they review) focus primarily on deterministic simulators. Many important problems have simulators that are stochastic (or are framed as stochastic), and performing reliability analysis on these simulators requires methods that handle this noise appropriately. \cite{GP:BayExpDesExtREsp22} considers a simulator heteroscedastic Gaussian noise and heteroscedastic non-Gaussian noise is considered in \cite{MS:SeqSampExEvent18,  WHB:MooringKriging, Gramstad:2020:Seq, WGSMVanem:compareULS24}.

This work presents a new method for estimating quantiles of extreme response distributions for black box stochastic simulators with non-Gaussian noise. We propose an estimation method that does not rely on long-term Monte Carlo simulation, which can improve estimation run time. This opens the door for acquisition functions that were previously prohibitively costly. The mathematical foundations of this approach are presented in Section \ref{sec:problem_formulation} and the required numeric approximations are introduced in Section \ref{sec:numerical_approximation}. Section \ref{sec:algo} combines these, presenting the full method for extreme response calculation and performing DOE. The expected benefits and limitations of the method are discussed in Section \ref{sec:discussion}.

\section{Problem formulation}
\label{sec:problem_formulation}
Here we introduce the necessary mathematical framework. We first define the extreme structural response $Y_{N_y}$, which is the distribution of the largest response a given structure will experience in a period of $N_y$ years. We then define a fixed quantity $z$ that depends on the distribution $Y_{N_y}$. For instance, $z$ could be the expectation or some percentile value of $Y_{N_y}$. Our main goal is to efficiently and accurately estimate $z$. To do this we will introduce a surrogate model for the structural response, which yields an approximation $\hat{z}$ of $z$. The Bayesian optimal experiment is the experiment that is expected to reduce the variance of $\hat{z}$ the most, and we define our Design of Experiments (DOE) strategy accordingly.

\subsection{The extreme response characterization}

We define random variables with upper case and fixed realizations with lower case. Variables in bold are vectors. 
We will make use of the following random variables:

\begin{itemize}
    \item The long-term environmental variable $\X$. $\X$ is a continuous random vector with joint probability density $f_\x(\x)$. We assume that one realization $\x$ of $\X$ parametrizes the short-term environment (represented as a stochastic process) for a period of length $T_s$. $T_s$ may be in the order of minutes or hours, depending on the environmental variables in question. In ocean engineering, the vector $\x$ typically contains variables that represent how the wind and waves (the environment) behave within a given time period (e.g. average wind speed, significant wave height, etc.).
    \item The short-term response $Y|\x$. For a \emph{fixed} long-term environmental realization $\x$, the maximum physical response within a time interval of $T_s$ is a random variable $Y|\x$ with probability density $g_{y|\x}(y|\x)$.  In an engineering context, randomness in $Y|\x$ typically comes from variability in the short-term environment conditions (e.g. randomness in the instantaneous conditions, which are generated by the stochastic process parametrised with the given $\x$), and from the variability in the response to identical short-term (instantaneous) conditions. We assume an accurate but expensive simulator $sim(\x)$ is available for sampling $Y|\x$.
    \item The marginal maximum physical response within a time interval of $T_s$ is denoted $Y$. For a \emph{random} long-term environment $\X$, $Y$ has density 
    \begin{equation*}
        g(y) = \int g_{y|\x}(y|\x) f_\x(\x) \ d\x.
    \end{equation*}
    \item The long-term $N_y$-year extreme structural response $Y_{N_y}$. If $T_s$ is defined in terms of hours, then let $N = \lceil N_y \cdot 365.25 \cdot 24 / T_s \rceil$. The $N_y$-year extreme response is defined as the maximum 
    \begin{equation*}
        Y_{N_y} = \max \{ Y_1, \dots, Y_N \}
    \end{equation*}
    where $Y_1, \dots, Y_N$ are i.i.d. with the same distribution as $Y$. In particular, if $Y$ has CDF 
    $$G(y) = \int_{-\inf}^{y} g(y') dy' ,$$ 
    then $Y_{N_y}$ has cdf
    $$ G_{N_y}(y) =  G(y)^N .$$\label{eq:extrapolation}
\end{itemize}

\subsection{Quantity of Interest}

Our ultimate focus lies in accurately estimating a statistic of the long-term extreme response distribution $Y_{N_y}$. This statistic is refereed to as the Quantity of Interest (QOI) $z$.  
In this paper, we will consider a quantity of interest of the form 
\begin{equation}
    \label{eq:qoi}
    z = G_{N_y}^{-1}(p),
\end{equation}
for a selected probability $p \in (0, 1)$. I.e. $z$ is a given percentile of the long-term extreme. 
However, our methodology is flexible and can be adapted to estimate various other relevant parameters as well.

\subsection{Surrogate modelling for Extreme Response Estimation}\label{sec:surrogate_modelling}
To address the computational demands of repeatedly simulating the distribution of $Y|\x$ for varying environmental inputs $\x$, we introduce a computationally efficient surrogate model, $\hat{Y}|\x$.  This is particularly relevant for applications such as calculating the bending responses of offshore wind turbine blades, where full aero-servo-elastic simulations can be computationally expensive due to complex non-linear physics and high dimensional inputs. 

For flexibility in approximating stochastic responses, we describe $\hat{Y}|\x$ as a member of a parametric family of distributions where the parameters are given by an unknown function $\theta(\x)$, similar to \cite{Gramstad:2020:Seq}. For example, $\hat{Y}|\x$ might follow a Weibull distribution with parameters determined by the functions $\theta(\x) = (\alpha(\x), \beta(\x))$.  We denote the probability density of this distribution as:

\begin{equation*}
    \hat{g}_{y|\theta}(y|\theta)
\end{equation*}

We use Gaussian process regression (GP) (see \cite{rasmussen2006gaussian}) to estimate the unknown function $\theta(\x)$ using a limited set of computationally intensive simulations. $\hat{\theta}(\x)$ is the GP estimator of $\theta(\x)$. Under general conditions (e.g. $\theta(\x)$ is continuous), the accuracy of this approximation can be improved with additional simulations. Using GP regression crucially allows us to:

\begin{itemize}
    \item Quantify the uncertainty in our estimation of $\theta(\x)$, i.e. the likelihood of different estimation errors $\hat{\theta}(\x_i) - \theta(\x_i)$.
    \item Evaluate how future experiments can reduce this uncertainty.
    \item Propagate this uncertainty to assess our confidence in the quantity of interest $z$.
\end{itemize}

\subsection{The data-generating process}\label{sec:data_generation}
Surrogate models require training data, which we generate using the simulator $sim(\x)$. We assume that the stochastic short-term response $Y|\x$ can be accurately modelled by the chosen parametric distribution family with parameters $\theta(\x)$ as described in Section \ref{sec:surrogate_modelling}. $sim(\x)$ provides samples of $Y|\x$ which can be used to estimate $\theta(\x)$. This is done using maximum likelihood estimation which estimates the mean of each parameter (i.e. the most likely parameter value given the samples) and the covariance of the mean estimate (i.e. using the Fisher information to compute the covariance matrix). The uncertainty in our estimate of $\theta(\x)$ is then asymptotically Gaussian, and our noisy observation $\theta_{obs}(\x)$ of $\theta(\x)$ generated by using the simulator can be expressed as:
\begin{equation}
    \label{eq:theta_obs}
  \theta_{obs}(\x) = \theta(\x) + \varepsilon(\x)  
\end{equation}

The observational noise, $\varepsilon(\x)$, is modelled as a multivariate Gaussian distribution with zero mean and covariance matrix $\Sigma(\x)$:

$$\varepsilon(\x) \sim N(\bm{0}, \Sigma(\x))$$

After $k$ simulations, we obtain a dataset:

$$D_k = \{ (\x_1, \bm{\theta}_{obs\_1}, \Sigma_1), \dots, (\x_k,\bm{\theta}_{obs\_k}, \Sigma_k) \}$$

\begin{remark}
    We could also let the estimate of $\theta(\x)$ depend on the number of samples, which is the number $sim(\x)$ evaluations for the given input $\x$, and in return find the optimal number of samples to generate for each experiment. However, for simplicity, we will consider $\x$ as the only input to the short-term response simulation here.
\end{remark}

\begin{remark}
A possible model simplification is to model the noise $\varepsilon$ as a Gaussian distribution with a diagonal covariance matrix $\Sigma$. I.e. that the components of $\theta(\x)$ are uncorrelated. 
\end{remark}

\subsection{Uncertainty modelling}

We identify two primary sources of uncertainty in data-generating process:

\begin{itemize}
    \item \textit{Epistemic Uncertainty:}  Our knowledge of the true $\theta(\x)$ is limited to the surrogate estimate $\hat{\theta}(\x)$. In the Gaussian process framework, we quantify the uncertainty in the estimate using a Gaussian random variable. Strategic selection of new experiments can reduce this uncertainty.
    \item \textit{Aleatory Uncertainty:} Due to observational noise $\varepsilon(\x)$, we cannot perfectly predict new observations even with perfect knowledge of $\theta(\x)$. This uncertainty is inherent to the simulation process and cannot be reduced through additional experiments.
\end{itemize}

Both of these sources of uncertainty will have to be taken into account when estimating the long-term extreme structural response.  

\subsection{The Gaussian Process regression model}

We model the uncertain relationship between environmental inputs $\x$ and the uncertain response parameter vector $\theta(\x)$ using a Gaussian Process (GP). After collecting $k$ observations ($D_k$), we obtain the conditional GP 
\begin{equation*}
\hat{\theta}_k(\x) = \hat{\theta}(\x) | D_k. 
\end{equation*}
As with standard GP regression, the mean and covariance of $\hat{\theta}_k(\x)$ can be computed analytically (see e.g. \cite{rasmussen2006gaussian}).
The GP framework provides a probabilistic interpretation, where we view $\hat{\theta}_k(\x)$ as a distribution over functions, defined on a probability space $(E, \mathcal{E}, P_k)$. Here, $E$ is a set, and a specific realization, $\hat{\theta}_k(\x| e)$ for a fixed $e \in E$, corresponds to a single function drawn from this distribution.

Crucially, by "freezing" a realization of the GP,  we can use the parametric surrogate model to simulate the response $Y|\x,e$. The probability density of this response is  $\hat{g}_{y|\theta}(y|\hat{\theta}_k(\x| e))$. To account for the variability in the long-term environment, we marginalize over the environmental distribution:

\begin{equation}
\label{eq:ghat_e}
\hat{g}(y|e) = \int \hat{g}_{y|\theta}(y|\hat{\theta}_k(\x| e)) f\x(\x) \ d\x
\end{equation}

This GP formulation provides a way to model the uncertainty in our estimation of $\theta(\x)$ and to propagate that uncertainty through the calculation of extreme responses. Additionally, ``freezing" a realization allows us to analyse the system behaviour and corresponding extreme events for a particular function within the GP's distribution.

\subsection{Design of Experiments}\label{sec:problem_formulation:doe}
Our primary goal is to minimize the uncertainty in estimating the quantity of interest $z$ with as few expensive experiments as possible. We achieve this by focusing on new experiments that contribute the most to reducing the variance of its estimate, $\hat{z}$.  This estimator is conditioned on a fixed realization $e$ of the Gaussian process and is defined as:
$$\hat{z} = \hat{z}(e).$$
With the quantity of interest in \eqref{eq:qoi}, $\hat{z}(e)$ will be the $p$-th percentile of the random variable $\hat{Y}_{N_y}|e$
$$\hat{z}(e) = \hat{G}_{N_y}^{-1}(p|e),$$
where $\hat{G}_{N_y}^{-1}(p|e)$ is the extreme value inverse CDF corresponding to the density $\hat{g}(y|e)$. 

We quantify the uncertainty about $\hat{z}$ using its variance, denoted as $H_k$, under the probability measure $P_k$ which is conditioned on the dataset $D_k$\footnote{The GP surrogate model $\hat{\theta}_k(x)$ serves as the probability measure $P_k$}:
\begin{equation}
    \label{eq:Hk}
    H_k = var_{P_k}[\hat{z}]
\end{equation}

The overarching objective of our Design of Experiments (DOE) strategy is to minimize $H_j$, where $j$ is the total budget of experiments that can be performed ($k \le j$). We do this by repeatedly picking the next data point which minimises $H_{k+1}$. Since $H_{k+1}$ is dependent on the outcome of the next experiment (which is unknown) we adopt a Bayesian approach and minimize its expectation. To fully account for all uncertainties, we must also consider the aleatory uncertainty stemming from the observational noise $\varepsilon(\x)$.  We represent the probability measure of this aleatory uncertainty as $Q_k$.

This leads us to the following acquisition function, which accounts for both epistemic and aleatory sources of uncertainty:
\begin{equation}
    \label{eq:acq}
  s_k(\x) = \mathbb{E}_{P_k \times Q_k}[H_{k+1}]  
\end{equation}

Our DOE strategy is based on starting with an initial dataset $D_k$ sparsely covering the input space. A common choice for this initial dataset is a space-filling design, which provides broad initial coverage of the input space.  Subsequently, we iteratively select the next experimental input which minimizes the acquisition function:

$$\x_{k+1} \in \argmin s_k(\x).$$

This strategy greedily guides our choices of subsequent experiments to perform, prioritizing those that are most likely to reduce the uncertainty in our estimate of the quantity of interest $z$. 

\section{Numerical approximation}
\label{sec:numerical_approximation}
Here we present some background on the approximations we will use in the algorithm for computing the acquisition function $s_k(\x)$ in \eqref{eq:acq}. 

\subsection{The need for approximation} \label{sec:need_for_approx}
If we want to implement the acquisition function \eqref{eq:acq} directly we will face the following challenges: 

\begin{enumerate}
    \item It is not possible to "freeze" a realization of a GP by fixing the variable $e$. We do not have the mapping from $e$ to the corresponding GP realization. Also, $e$ is infinite-dimensional. Due to the memory demands, it is also impractical to compute the posterior over all inputs and sample a "frozen" GP realization from said posterior.
    \item Evaluating the integral in Eq. \ref{eq:ghat_e} with crude Monte Carlo simulation will be inefficient as a large number of samples are needed to estimate rare $\X$ events.
    \item Evaluating $H_k$ in \eqref{eq:Hk} and $s(\x)$ \eqref{eq:acq} with crude Monte Carlo simulation will be inefficient as it requires nested Monte Carlo with a non-trivial calculation \eqref{eq:ghat_e}. 
    \item When the observational noise $\varepsilon(\x)$ in \eqref{eq:theta_obs} depends on $\x$, we do not have a full generative model. I.e., we cannot simulate future observations without knowing $\Sigma(\x)$.
\end{enumerate}

\subsection{Finite-dimensional GP approximation and the unscented transform}
We make use of two techniques developed in \cite{Agrell:2020:DOE_SRA} to obtain a numerically efficient approximation of the acquisition function. We use a finite-dimensional approximation of the GP, and then use the unscented transform on this approximation to perform efficient uncertainty propagation. 

\subsubsection{Finite-dimensional GP approximation}
Let $\hat{\theta}(\x)$ be a multivariate Gaussian random variable. Then 
\begin{equation}
    \label{eq:std_norm}
    \hat{\theta}(\x) = E[\hat{\theta}(\x)] + L(\x)\UU,
\end{equation}
where $\UU$ is a multivariate standard normal variable, $\UU \sim N(\bm{0}, I)$, and $\text{cov}[\hat{\theta}(\x)] = L(\x)L(\x)^T$. To approximate a single "frozen" realization of the GP (i.e. $\hat g(y|e)$) a single sample of $\UU$ is used for each point-wise prediction given by the GP. This approximation means that for two inputs $\x$ and $\x'$, the function values $\hat{\theta}(\x)$ and $\hat{\theta}(\x')$ will be fully correlated. See Fig. 2 in \cite{Agrell:2020:DOE_SRA} for an illustration of what this approximation looks like in practice. This approximation addresses point 1 in Section \ref{sec:need_for_approx}.

\subsubsection{The unscented transform (UT)}
Let $\UU$ be a multivariate standard normal variable and $f$ a non-linear function. The unscented transform is a very efficient method for approximating the mean and covariance of the transformed variable $f(\UU)$.

We start by defining a set of sigma-points for $\UU$. This is a set of weighted samples $\{ (v_1, \uu_1),$ $\dots, (v_n, \uu_n) \}$. See Appendix B.2 in \cite{Agrell:2020:DOE_SRA} for how to compute these.

The UT approximation of the mean and covariance of $\bm{F} = f(\UU)$ are then obtained as 
\begin{equation}
    \label{eq:UT_moments}
    \begin{split}
        &\widehat{E}[\bm{F}] = \sum_{i=1}^{n} v_i \f_i, \\ 
        &\widehat{\text{Cov}}[\bm{F}] = \sum_{i=1}^{n} v_i (\f_i - \widehat{E}[\bm{F}])(\f_i - \widehat{E}[\bm{F}])^T,
    \end{split}
\end{equation}
where $\f_i = f(\uu_i)$.

The UT approximation is efficient as $n$ is typically small. In our implementation, we will use a method for computing the sigma-points where $n = 2\cdot \text{dim}(\bm{F}) + 1$. This approximation addresses point 3 in Section \ref{sec:need_for_approx}.

\subsection{Importance sampling}\label{sec:numerical_approximation:importance_sampling}

Let $h_{\x}(\x)$ be a probability density with the essential property that it is non-zero only where the original environmental probability density, $f_{\x}(\x)$, is also non-zero. This allows us to rewrite the marginal response density:
\begin{align*}
    g(y) &= \int g_{y|\x}(y|\x) f_\x(\x) \ d\x \\ 
    &= \int g_{y|\x}(y|\x) \frac{f_\x(\x)}{h_\x(\x)} h_\x(\x) \ d\x \\
    &= \mathbb{E}_{h_\x} \left[ g_{y|\x}(y|\x)\frac{f_\x(\x)}{h_\x(\x)} \right] \\ 
    & \approx \frac{1}{M} \sum_{m=1}^{M} g_{y|\x}(y|\x_m)\frac{f_\x(\x_m)}{h_\x(\x_m)}
\end{align*}

where $\x_1, \dots, \x_M$ are iid samples from the distribution with density $h_{\x}(\x)$. Note, to ensure the approximation is a valid distribution the sample importance weights ($w_m = \frac{f_\x(\x_m)}{h_\x(\x_m)}$) are first linearly rescaled to ensure $1 = \sum_{m=1}^{M} w_m$. We can write: 

$$G(y) = \int_{-\inf}^{y} g(y') dy' \approx \frac{1}{M} \sum_{m=1}^{M} G_{y|\x}(y|\x_m)\frac{f_\x(\x_m)}{h_\x(\x_m)}.$$ 
Consequently, we may adopt the following approximation for the long-term extreme response distribution:

$$ G_{N_y}(y) \approx  \left( \frac{1}{M} \sum_{m=1}^{M} G_{y|\x}(y|\x_m)\frac{f_\x(\x_m)}{h_\x(\x_m)} \right)^N.$$
Finally, we can replace the unknown $G_{y|\x}$ with its analytical counterpart from the surrogate model, $\hat{G}(y|\theta(\x))$. This approximation addresses point 2 in Section \ref{sec:need_for_approx}.

\subsection{Predicting the observation noise for new inputs \texorpdfstring{$\x$}{x}}
\label{sec:likelihood_approx}

Recall from Section \ref{sec:problem_formulation:doe} that $H_{k+1}$ is computed by adding the data point $(\x_{k+1}, \bm{\theta}_{obs\_k+1}, \Sigma_{k+1})$ to $D_k$. As $\bm{\theta}_{obs\_k+1}$ and $\Sigma_{k+1}$ are unknown (the simulator needs to be run to get them), we calculate the expected $H_{k+1}$ (see Eq. \ref{eq:acq}). This requires a distribution over possible $\bm{\theta}_{k+1}$ and $\Sigma_{k+1}$ values. The GP can be used to estimate the $\bm{\theta}_{k+1}$ distribution, but no prediction is available for $\Sigma_{k+1}$. While $\Sigma_{k+1}$ could be comprehensively modelled with an additional GP, we believe for some problems simple deterministic heuristics are enough as this parameter is often not exceedingly influential. A GP or heuristic can be used to address point 4 in Section \ref{sec:need_for_approx}.

We introduce the heuristic:
$$\tilde{\Sigma}(\x) \approx \Sigma(\x).$$
We will let $\tilde{\Sigma}(\x) = \Sigma_j$ where $j$ refers to the experiment in $D_k$ most similar to $\x$ (e.g. $\x_j$ is the one closest to $\x$ in Euclidean norm).

\section{Algorithm}\label{sec:algo}
Here we present an algorithm for implementing the DOE method described in \secref{sec:problem_formulation} using the practical approximations introduced in \secref{sec:numerical_approximation}.

For clarity, we explicitly state the following assumption regarding the Quantity of Interest (QOI): Our algorithm will focus on a specific percentile of the extreme value distribution $G_{N_y}(y)$. We denote our QOI as
$$ z = G_{N_y}^{-1}(p),$$
where $p$ represents a chosen quantile (e.g., 0.5 for the median).

\subsection{Initialisation}\label{sec:algo:init}

Our algorithm begins with the following initialization steps:
\begin{enumerate}
    \item \textbf{Importance Sample Generation:}  We generate a set $\{ \x_{m} \}_{m = 1, \dots, M}$ where each sample $\x_{m}$ is drawn from the importance sampling distribution $h_{\x}(\x)$. This distribution should be chosen to cover input regions that significantly impact our extreme response estimates.  See Appendix \ref{app:importance_sampling} for a suggested sampling distribution. We then calculate importance sampling weights, $w_m = f_{\x}(\x_m) / h_{\x}(\x_m)$. These weights adjust for the fact that we are sampling from $h_{\x}(\x)$ and not directly from our target density $f_{\x}(\x)$. 
    \item \textbf{Unscented Transform Preparation:} We compute sigma points  $\{ (v_1, \uu_1), \dots, (v_n, \uu_r) \}$ for a multivariate standard normal variable. The equations in Appendix B.2 in \cite{Agrell:2020:DOE_SRA} can be used to compute sigma-points. With this method we will have $r = 2\cdot \text{dim}(\theta) + 1$.
\end{enumerate}

\subsection{Algorithm for computing \texorpdfstring{$H_k$}{Hk}}
\label{sec:H_k}

The core steps for computing $H_k$, our measure of uncertainty in the estimated quantity of interest, are:

\begin{enumerate}
    \item \textbf{Gaussian Process (GP) Evaluation:} For each sample $\x_m$, we obtain the following from our Gaussian Process surrogate model:
    \begin{itemize}
        \item $\mu_{m} = E[\hat{\theta}(\x_{m})]$ : The expected value of the parameter vector at $\x_m$.
        \item $\Sigma_{m} = \text{Cov}(\hat{\theta}(\x_{m}))$: The covariance matrix, capturing uncertainty in $\theta(\x_m)$.
        \item $L_m$: The Cholesky factor of $\Sigma_m$. Alternatively, any matrix $L_{m}$ such that $L_{m}L_{m}^T = \Sigma_{m}$ will work.
    \end{itemize}
    \item \textbf{Unscented Transform for $\theta$:} We generate a set of weighted samples to approximate the distribution of $\theta(x_m)$ using the UT:
    $$\btheta_{m, q} = \mu_{m} + L_{m} \uu_q \quad \text{for } q = 1, \dots, r$$
    \item \textbf{Sample $\hat{z}$: } We leverage both importance sampling and the analytical distribution for short-term response $\hat{G}$:
    \begin{itemize}
        \item We compute the value of our QOI under each UT realization of $\theta$:
      $$\hat{z}_{q} = \hat{G}_{N_y, q}^{-1}(p)$$
      \item The long-term distribution is efficiently approximated:
      $$\hat{G}_{N_y, q}(y) = \left( \frac{1}{M} \sum_{m=1}^{M} \hat{G}(y | \btheta_{m, q}) w_m \right)^N$$
       Note the use of importance sampling weights $w_m$ here. Recall also that $N = \lceil 365.25 \cdot 24 / T_s \rceil$ is the (large) number of short-term time-periods within $N_y$ years. But we will use $M<<N$ samples. And $\hat{G}(y | \btheta_{m, q})$ is the known CDF from the family we have chosen (e.g. Weibull). We will have to invert $\hat{G}_{N_y, q}(p)$ in total $r$ times, so the fact that $r$ is small is important. 
    \end{itemize}
    \item \textbf{Calculating $H_k$:} Finally, we obtain our uncertainty measure using the UT approximations for expectation and variance:
    \begin{itemize}
        \item UT approximation of $E[\hat{z}]$:
        $$\hat{\mu}_{\hat{z}} = \sum_{q = 1}^{r} v_q \hat{z}_{q} $$
        \item UT approximation of $H_k = var[\hat{z}]$:
        $$\hat{H}_k = \sum_{q = 1}^{r} v_q (\hat{z}_{q} - \hat{\mu}_{\hat{z}})^2$$
    \end{itemize}
\end{enumerate}

\subsection{Algorithm for Computing \texorpdfstring{$s_k(\x)$}{sk(x)}}

The computation of $s_k(\x)$ is the basis for the experimental design strategy. Recall that $s_k(\x)$ quantifies the expected uncertainty about our quantity of interest $z$, if we include an additional experiment with input $\x$. It can be calculated as follows:

\begin{enumerate}
    \item \textbf{GP Evaluation at Input $\x$:}  We begin by obtaining the relevant statistics from the Gaussian process surrogate model:
    \begin{itemize}
        \item $\mu = E[\hat{\theta}(\x)]$, 
        \item $\Sigma = \text{Cov}(\hat{\theta}(\x))$, and 
        \item $L$, the Cholesky factor of $\Sigma$
    \end{itemize}
    \item  \textbf{Unscented Transform for $\theta$:}  We generate sigma points representing the distribution of  $\theta(\x)$:
    $$\btheta_{q} = \mu + L \uu_q \quad \text{for } q = 1, \dots, r$$
    \item \textbf{Uncertainty Propagation for $H_{k+1}$:} For each sigma point $\btheta_q$,  we simulate the effect of observing this potential outcome at input $\x$:
    \begin{itemize}
        \item \textbf{GP Update:}  Create a hypothetical dataset  $D_{k+1} = D_k \cup \{(\x, \btheta_{q}, \tilde{\Sigma}(\x))\}$, and use this to update the GP (See Section \ref{sec:likelihood_approx} for details on $\tilde{\Sigma}(\x)$). This gives  a new GP,  $\hat{\theta}_{k+1}$, reflecting the information we would have if we observed $\btheta_q$ at $\x$
        \item \textbf{Compute  $H_{k+1}$:}  Using the updated GP $\hat{\theta}_{k+1}$, compute $H_{k+1}$ following the algorithm from Section \ref{sec:H_k} (the $H_k$ algorithm). Store this value as $\hat{H}_{k+1, q}$.
    \end{itemize}
    \item \textbf{Compute $s_k$:} 
    $$ \hat{s}_k(\x) =  \sum_{q = 1}^{r} v_q \hat{H}_{k+1, q}.$$
\end{enumerate}

\section{Discussion}\label{sec:discussion}

In this paper we presented the mathematical foundation for a new method of calculating quantiles of the extreme response distribution using a Gaussian process surrogate model. We also present several numerical approximations used to make this method more efficient. These efficiencies allow for new DOE approaches.

While surrogate models are significantly faster than real simulators, the large number of surrogate runs required to perform a Monte Carlo estimate can still be prohibitive. Assuming a simulation time steps of 1 hour, a single Monte Carlo estimate of the 25 year extreme response would require $\approx$220,000 surrogate runs. To estimate the quantity of interest associated with the extreme response distribution, this process would need to be repeated many times. The $M$ surrogate runs required by the proposed method is much lower (for example $\approx$5,000 might be suitable depending on dimensionality). An additional benefit is that $M$ is independent of $N_Y$ (the number of surrogate time steps in the time period of interest). As such, the proposed method scales better to longer time frames or more granular simulators.

The computational improvements facilitate a range of Design of Experiments (DOE) methods that might be prohibitively expensive using crude Monte Carlo estimation. We hypothesise the mathematically founded DOE outlined in this paper may offer efficiency improvements. It may also be more robust and generalizable than domain-specific heuristic DOE methods.

A limitation of this method is that we must approximate the marginalization $G(y)$ in Eq.\ref{eq:extrapolation}, and this approximation then has a large exponent applied. As a result, small errors in the approximation of $G(y)$ will compound into larger ones. Similar to Monte Carlo methods, the choice and fit of distribution used by the surrogate (Section \ref{sec:surrogate_modelling}) can have a significant impact on the result. In particular, the fit quality in the tail regions needs to be carefully considered. Quantifying the impact of these issues should be further explored in future work.

The next stage of this work should demonstrate the method's performance empirically. A test problem should be found that is reflective of real-world problems, while still computationally efficient enough to produce brute force Monte Carlo results.

This mathematically founded method offers a domain-agnostic approach for efficiently estimating extreme response quantiles, and systematically reducing the uncertainty in these estimates. This means the method can be applied across a wide range of industries and problems, including those where good existing domain-specific heuristics are not available. While this method was developed primarily for ULS calculations in engineering, it is broadly applicable to problems that require understanding the long-term behaviour of a slow, stochastic system affected by uncertain inputs. Better DOE methods lead to better estimates with the same simulation budget. With better estimates uncertainty is reduced, allowing decision makers to make more precise, uncertainty-informed choices.

\section{Summary and Conclusion}

In this paper we present a method for estimating the long-term extreme behaviour of stochastic black box models operating under uncertain conditions. The proposed method captures the impacts of variability in both the long-term environment and the short-term response. It also offers significant efficiency gains compared to similar Monte Carlo approaches. Due to its general and well-founded formulation, we speculate this approach might be more performant and widely applicable than domain-specific Design of Experiment heuristics. This promising method needs to be further validated with empirical results. A generalizable method has the potential to make extreme response estimation easier across a range of industries, reducing the uncertainty decision-makers face and leading to more informed choices and better outcomes.

\bibliographystyle{plain}
\bibliography{bibliography.bib}

\appendix
\section{Selecting the importance sampling distribution \texorpdfstring{$h_{\x}(\x)$}{h(x)}} \label{app:importance_sampling}
Importance sampling is a technique used to improve the efficiency of our extreme response estimations, as detailed in Section \ref{sec:numerical_approximation:importance_sampling}. Importance sampling requires that we design a sampling distribution $h_{\x}(\x)$ that focuses on the regions of the input space most likely to influence extreme events. An effective $h_{\x}(\x)$ for extreme response analysis should have the following properties: 

\begin{itemize}
    \item \textit{Focus on Extremes:} It should assign a high probability to the extreme environmental conditions ("storms") that are critical for accurately modelling the tail behaviour of the extreme response distribution.
    \item \textit{Exploration of Typical States:} Simultaneously, the distribution should also adequately sample typical (or 'normal') environmental states, as these provide essential context for the system's general behaviour.
\end{itemize}

\textbf{Proposed Strategy:} We propose a strategy based on uniform sampling within a selected region:

\begin{enumerate}
    \item \textit{Define the Region:}  Introduce a threshold $c$ and define the set: 
    $$F = \{\x | f_{\x}(\x) > c \}$$
   This set will be the "practical support" of the density $f_{\x}$. I.e. the threshold $c$ should be set so that we can assume any realization outside $F$ effectively has probability zero. 
   \item \textit{Sample Generation:}
   We define $h_{\x}(\x)$ as the distribution that is uniform on $F$. We can sample from this distribution as follows.
   \begin{itemize}
       \item Create a hyper-rectangle $V$ such that $F \subseteq V$. Let $|V|$ denote the volume of the hyper-rectangle.
       \item Generate $M_\text{tot}$ samples uniformly on $V$.
       \item Discard any samples that fall outside the set $F$, and let $M$ be the number of remaining samples (i.e. where $f_{\x}(\x_i) > c$). The remaining samples are i.i.d. samples from $h_{\x}(\x)$.
   \end{itemize}
   \item Density Estimation: The density of $h_{\x}(\x)$ is constant and can be estimated as:
   $$h_{\x}(\x) \equiv \frac{1}{|F|} \approx \frac{M_{\text{tot}}}{|V|M}$$
\end{enumerate}

One key point here is that the threshold $c$ provides flexibility in adjusting the balance between focusing on extreme scenarios and maintaining the representation of typical environmental conditions.

\end{document}